# Toward Developing Intelligent Mobile Obe System In Higher Learning Institution

abdifatah Farah Ali, rusli Haji Abdulah, mohamed M. Mohamed

**Abstract:** The rapid growth in Mobile application users has made the researchers and practitioners to think of intelligent tools that can help the users and applications in delivering quality of services. Therefore, intelligent agent is expected to become the tool for development of mobile outcome based education (OBE) particularly in higher learning Institutions (HLI). In this context, there is a lacking of OBE intelligent agent in assisting the academicians to use in OBE management for mobile application system environment. This paper presents the conceptual design and development of a mobile intelligent agent based on mobile OBE called as i-MOBE. Since that, i-MOBE that we developed is considered very important for academicians and students to facilitate them in using for academic purpose in HLI particularly in helping them to monitor the performance in teaching and learning (T&L). The system architecture will be covering the conceptual design and its interaction as well as the system configuration in helping the academicians to use the system in their T&L toward effective and efficiency also can be applied in monitoring based on scenarios such as test, assignment and projects and so on.

**Keywords:** outcome based education; mobile application; mobile agent; higher learning institution; teching and learning.

——————————◆——————————

## 1. INTRODUCTION
The flexibility, programmability, and extensibility (accepting significant extensions or amendments of functionalities without a fresh re-writing of code or gross change in the architecture), are major things found lacking in today's Internet systems, especially for supporting applications and other new services which are interconnect on separate or distinct networks. Hence we need to introduce intelligent tools to overcome the problem efficiently. The subject of software agent being a new technology paradigm in the areas such as artificial intelligence and computing system, they facilitates these features in Internet service and application development [1-3]. It is projected that agents will dominate next generation components in software due to its inherent behavior and structure supporting CBSE (Component based software engineering) [4, 5]. Agents are programs, which are autonomous and are situated within an environment; they achieve their goals actively by sensing the entire environment. They use internal world state information and also inference engine in computing the action to be performed on the environment by sensing the reception messages from other users or the environment. [9]. they also have special characteristics, which differentiate them from standard programs: mandatory and optional properties. The orthogonal properties of the agents, provides a strong notion of the agents in turn [9, 10]. These properties are enumerated below:
- Autonomy: Operation of agents is done without direct intervention of humans or otherwise, with some kind control on their actions and their internal state.
- Decision Making: Here we have it in two perspective, first is the reactive decision making which is implemented by direct mapping from sensors input (sensing the environment) to actions by using some rules. While in proactive decision-making, can use BDI architecture (Belief Desire Intension) [11]. The facts of an agent are the model of its domain, its desires provides sort of ordering between states, and its intensions are things it decided to do. The intensions are defined at various levels of abstraction.
- Temporal continuity: Here agents are continuously running processes by either running active in the foreground or passive/sleeping in the background.
- Goal oriented: In this case, the agent is capable of handling a task in meeting its desired goal.

This paper is organized as follows: The background knowledge and related work will be described in Section II. In Section III, the research methodology is discussed related to the intelligent agent in OBE research. In Section IV, is the overview of proposed of conceptual design and its implementation. The scenarios based implementation from academician and student prospective is presented in Section V. In Section VI, is the data analysis for system evaluation and discussion. This paper is concluded in Section VII.

## 2. LETRAURE REVIEW
*A.* **Outcome based education (OBE)**
Outcome based education approach (OBE) has implemented or adopted by a number of countries such as Australia, Malaysia or USA for many years. OBE can be defined as "a comprehensive approach to organizing and operating an education system that is focused in and defined by the successful demonstration of learning sought from each student"[12]. OBE is an education approach that focuses on the learning outcome after completing an academic programme [13, 14]. OBE is a process that involves evaluation and assessment process in education to reflect the attainment of expected teaching and learning [15]. The process can be done through assessment method which consist of tests, assignments, presentations and projects. The development of learning outcome following the hierarchy of unit outcomes, lesson outcome, course outcome, exit outcome and program outcome especially in HIL [16].

*B.* **Mobile application**
A mobile application is a software application development especially for use on small computing devices such as tablets, smartphones rather than computers or laptops. These devices could be used to make phone calls, multimedia messages and send texts. These devices have the ability to connect people via wireless networks also has many features that help peoples to organize their time and resources and make the mobile phone suitable channels for educating, learning, development and training. Therefore, using mobile

________________________
- fitaaxfarah@simad.edu.so, rusli@upm.edu.my myare81@simad.edu.so
- 1, Faculty of Computing, SIMAD UNIVERSITY, Mogadishu, Somalia
- 2, Department of Software Engineering and information system,University Putra Malaysia, Selangor, Malaysia
- 3Faculty of Engineering, SIMAD UNIVERSITY, Mogadishu, Somalia

phone in HLI is an innovative opportunity that offers existing learning channels [17]. Hence, mobility is seen by researchers as a new opportunity for education and it provides for learners to personalize their learning process, enhance collaboration and social interactions with others at anytime and anywhere in the learning context [18].

*C.* **Mobile Agent**
It's amazing that Mobile agents (MA's) are found to be exciting and as a new programming paradigm, which are applied in many domains of computing systems like: information filtering, mobile computing, network management systems and in electronic commerce system, they are found to have many advantages in all the aforementioned domains. Systems implementing mobile agents have been in existence for quite some years, platforms supporting mobile agents are now available from industries and other research communities [19-21]. However, production environments applications are lacking the paradigm and it has acquired the reputation of being "a solution looking for a problem" [22]. Grossly, the delivery of the technology could not meet the initial deliverables expected by researchers or research environment. Mobile agents are independent programs having ability to move from one computer system to the other in a given network, and to other components/locations of their choice. State of running programs is saved in such that they are transmitted to the destination. The program process is resumed at the destination from the saved state. They are capable of providing an efficient, convenient and robust framework for implementation of distributed applications and as well gives smart environment for several reasons, like: latency and bandwidth improvements of client server systems and also reduction of vulnerability to network disconnection.
As a matter of fact, these mobile agents posses numerous advantages in developing various services at smart environment as well not just only distributed application systems.

## 3. METHODOLOGY
In order to formulate and propose the intelligent agent architecture of OBE system implementation for HLI as system architecture and its application, there are few steps that has been taken and conducted based on a series of sequence as shown in Figure 1.

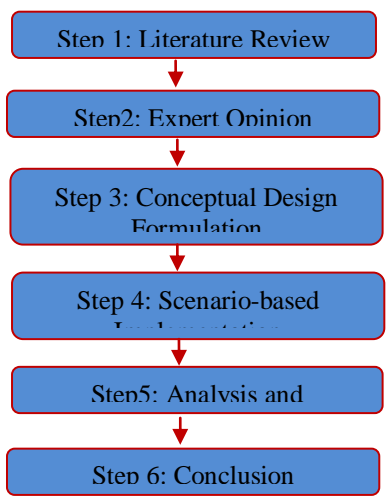

*Figure 1:* Research Design approach

The methodology of the research is started by performing the analysis of literature review (step1) regarding on OBE, mobile application and intelligent agent, then followed by conducting a preliminary survey (step2) through the expert opinion interviewed that based on those who are really involved in HLI such as academicians (Teachers) and students. At this stage, i-MOBE system for HLI as proposed conceptual system design and its architecture which closed to OBE system development (step3). After that, we described for the scenario-based implementation from two types of uses based on proposed platform (step4), then followed by the simple measurement OBE for mobile application that comprises three sections which are general information, user adoptive evaluation and system's capability as a conceptual system and its application system design model is also analyzed (step5) in determining the best criteria of Mobile Application for Android environment (step6) as an ending stage which also including the conclusion stage.

## 4. A PROPOSED CONCEPTUAL SYSTEM DESIGN

*D.* **The OBE Intelligent Paradigm**
The OBE intelligent paradigm consists of four agents: academic interface agent, student interface agent, assessment agent and database agent as shown in Figure 2. The overall architecture consists of three high level modules:

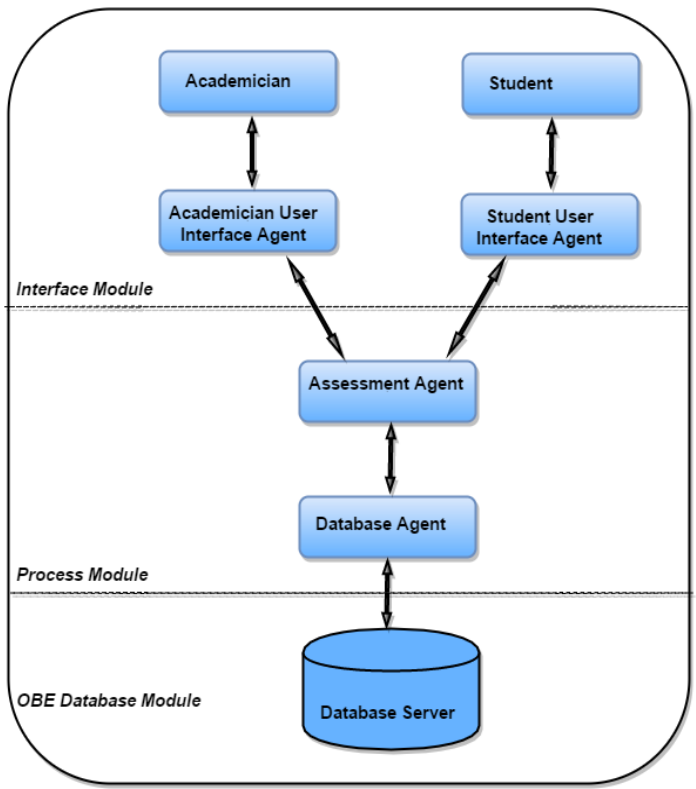

*Figure 2:* OBE Intelligent Paradigm Systems

1. **Interface Module :**

The interface module deals with the academic and student agent that are publicly visible. It provide mechnisms for interacting with agent and supports inter-agent collaboration and communication.

**2. Process Module :**

The process module is restricted only to the agent that is directly manipulating the contents of the module with access privilege. This module contains methods that implement a variety of functions and processes using which agent can respond to request from other agents and also provide the services that may necessary in solving particular problem.

**3. OBE Database Module :**

The OBE Database Module is restricted only to the agent that is directly manipulating the contents of the module with access privilege. This domain provide domain specific relevant to problem solving. It's responsible for keeping track of what data are stores in the main database and also responsible for retrieving the necessary data requested by assessment agent through database agent.

*E.* **The i-MOBE System architecture**

The i-MOBE system architecture consist of six architectural features: Academic Agent (AA), Student Agent (SA), User interface Agent (UIA), System Administrator Agent (SAA), Assessment Agent (AssA) and OBE database server. The communications between agent collaboration are shown in Figure 3. The functions of each agent are described in Table 1.

*Table 1: Agent Functions and its accessibility*

| No | Agent/ accessibility | Functionality |
|---|---|---|
| 1 | Academic Agent (Public) | The Academic Agent (AA) is used by the academician (Teachers), enabling them in managing and evaluating a wide variety of assessment of their courses. In this unit, the academician can use their mobile devices to access the information or any other aspect of instructions at anytime and anywhere. For an Academic Agent to be activated, it must be dispatched to a specific data container. The network authenticates the Academic Agent by checking its credentials and determines its privileges and then assigns a thread to execute the Academic Agent which is sent to the Internet through an access point (AP). After the agent is connected to the internet, it will continue to perform its task. |
| 2 | Student Agent (Public) | The Student Agent (SA) is used by the students, enabling them in seeing a wide variety of assessment of their courses. In this unit, the students can use their mobile devices to access the information provided by the academician at anytime and anywhere. For Student Agent to be activated, it must be dispatched to a specific data container. The network authenticates the Student Agent by checking its credentials and determines its privileges and then assigns a thread to execute the Student Agent which is sent to the Internet through an access point (AP). After the agent is connected to the internet, it will continue to perform its task. |
| 3 | User Interface Agent (Public) | The User Interface Agent is a type of agent which serves as the user's gateway to access the network of MOBE agents. It's responsible for managing the user's request to the appropriate agents and given feedbacks to the user. The UIA is autonomous and can maintain the user's context beyond a single session and allowing long term autonomous data retrieval and other tasks to continue in the user's absence. The UIA receives the requests from the user interface application to invoke internal services. Therefore, the UIA is an access control mechanism which authenticates users before starting a client container also responsible for final presentation of results through a mobile agent before passing data to the application layer. |
| 4 | System Administraor Agent (Public) | The System Administrator Agent (SAA) is a type of agent which analyzes data from multiple users. The SAA is responsible for managing, maintaining and overseeing a multiuser computing environment by setting up users account and also performing procedures to prevent the spread of such viruses. |
| 4 | Assessment Agent (Private) | The Assessment Agent (AssA) is a type of agent which can be able to assess the functions of Academic Agent with respect to the User Interface Agent through the main container and it also can access to assess the Student Agent activity via User Interface Agent. |
| 5 | OBE database server (Pulic or Private) | The OBE database server is the core component of the system which is used for storing all data of the system and provides interfaces for the client to access it. The OBE database server consists of two information repositories which are data and user profiles. The Academic Agent collects data from a user (student) to Assessment Agent that subsequently stores the data in the repository. After that, a copy of data along with the user's profile is sent to Administrator Agent for real time monitoring. |

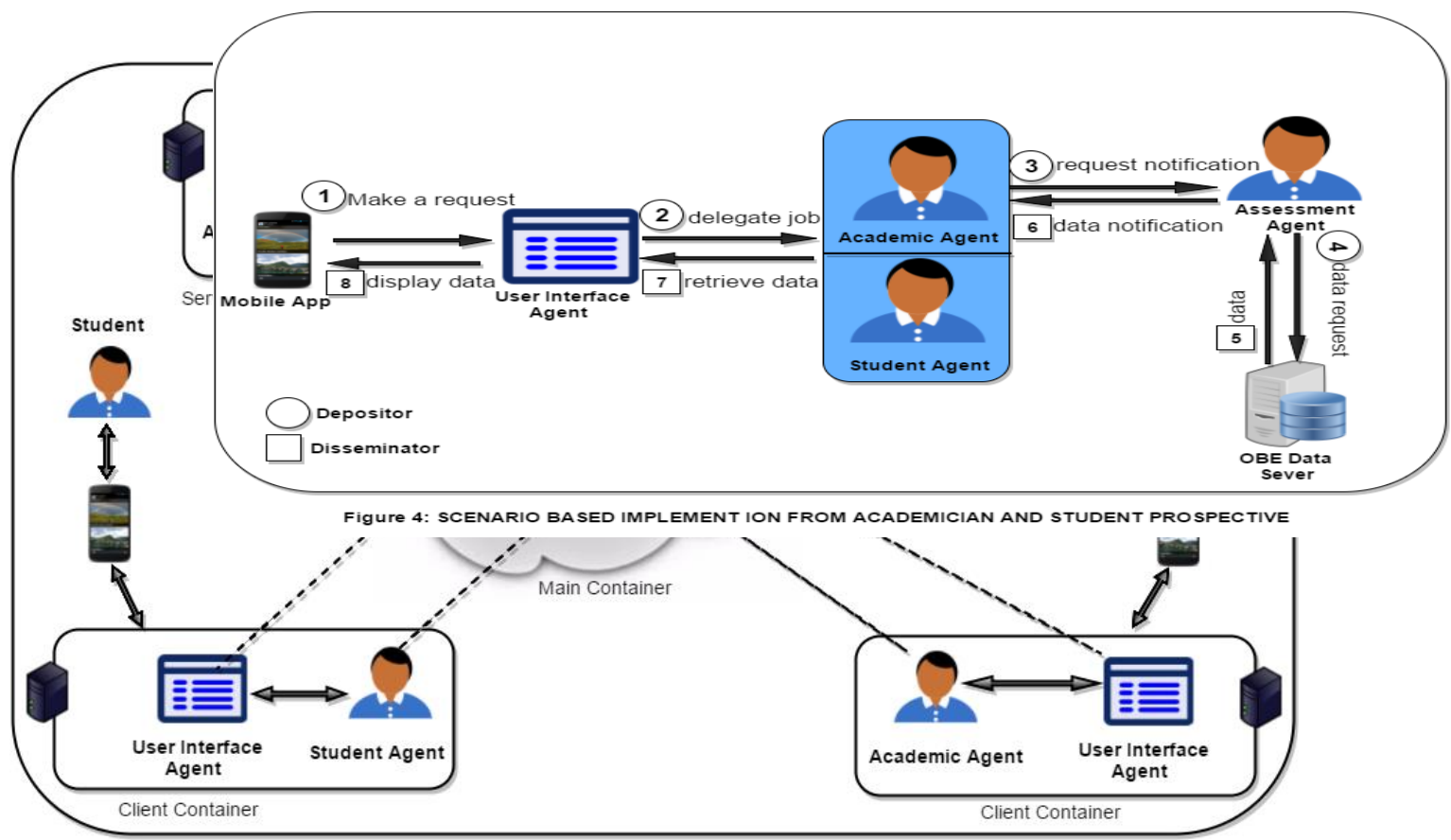

Figure 4: SCENARIO BASED IMPLEMENTION FROM ACADEMICIAN AND STUDENT PROSPECTIVE

Figure 3: i-MOBE SYSTEM ARCHITECTURE

## 5. SCENARIOS BASED I-MOBE IMPLEMENTATION

The proposed intelligent system architecture distinguishes two types of user: Academician (Teacher) and Student. The academician wants to perform a routine assessment on his student through a user interface on his mobile application and he/she request the UIA to collect his/her student's information. Upon receiving the request, the UIA launches and delegates the job to the Academic Agent. In the meantime, the Academic Agent surfs the logical networks to acquire the needed information. When the needed information found, the Academic agent asks or requests the assessment agent to retrieve the data. The Academic Agent continues delivering messages until it has searched all relevant information. After receiving the request notification, the Assessment Agent sends a query to the OBE database server for the relevant data. The OBE database server sends the data back to the Assessment Agent. After that, the assessment Agent sends the result back to the academic Agent. When the Academic Agent has collected all the required data, it returns to the client container which created it and sends the data to UIA. Finally, the UIA displays the data on the User interface of mobile Application. The process of scenarios based implementation as shown in the Figure 4.

## 6. DATA ANALYSIS AND DISCUSSION

This study was conducted at the University Purta Malaysia (UPM). A survey instrument was developed based on previous literature to determine the user's on the usability aspect of i-MOBE system. A total 100 questionnaires was distributed and collected based on 5 point likert scale ranging from 1 being strongly disagree, 2 = disagree, 3 = neutral (neither disagree nor agree), 4 = agree, 5 = strongly agree. The survey consisted of three parts. In the first part, the respondents were asked questions related to demographic information such as gender, education and so on. In the second part, the questions related to the user adaptive evaluation by investigating user perceptions of i-MOBE, the evaluation comprise five constructs: Agent Usefulness, Agent Ease of Use, Agent Information Quality, Agent save time and Agent efficiency. The third part, the respondents were asked questions related to the system's capability. The data was analyzed using IBM SPSS v 23. The participant in this study was voluntary. To make sure reliable content and increase content validity, we identified that the survey should be filled out by the respondent who is familiar with similar mobile application based on their usage experience and to reduce self reporting bias, we told all participants to receive the findings of the study.

### F. Profile of respondent

A sample of 100 randomly selected respondents, as shown in the Table 2, 63% of respondents were male and 37% were female, 30% of respondents were aged between 31 and 35 years old. Those in Sciences made up the largest group of respondents (44%), those in Business studies were 32% and lastly those in art studies were only 24%. In terms of education level, most of respondents were in Master level (47%), 82% of the participant declared that they own smart phone and 18 % own a simple Hand phone. Regarding the use of mobile application, 28% have less than 5 years of experience; 54% have experience between 5 and 10 years of use. Thus, it indicates that the respondents are familiar with the subject matter and suitable for this study.

*Table 2: Demographic Data*

| DISTRIBUTION | | FREQUENCY | Percentage (%) |
|---|---|---|---|
| Gender | Male | 63 | 63 |
| | Female | 37 | 37 |
| Age | Below 20 | 0 | 0 |
| | 20-25 | 11 | 11 |
| | 26-30 | 26 | 26 |
| | 31-35 | 30 | 30 |
| | 36-40 | 21 | 21 |
| | More than 40 | 12 | 12 |
| Area | Science | 44 | 44 |
| | Business | 32 | 32 |
| | Art Studies | 24 | 24 |
| Education Level | Bachelor | 30 | 30 |
| | Master | 47 | 47 |
| | PHD | 23 | 23 |
| Mobile Devices | Smart Phone | 82 | 82 |
| | Hand Phone | 18 | 18 |
| Mobile Application Experience | < 5 Years | 28 | 28 |
| | 5-10 | 54 | 54 |
| | >=10 years | 18 | 18 |

### G. User adoptive evaluation

The instrument dimensions of adaptive user evaluation as shown in Figure 5 and Figure 6, compares by their sections and the participants have given the Agent efficiency 94% that is good signed and also the Agent quality information provided by the system around 85% as well as accepts to be used the system in their environment. In the context of the Agent ease of use of mobile application they were given around 93%, also for the Agent usefulness (89%) and that is shows the participants accept of the mobile application implementation and their performing well to apply it in their teaching and learning environment.

*Figure 5: Adoptive user evaluation*

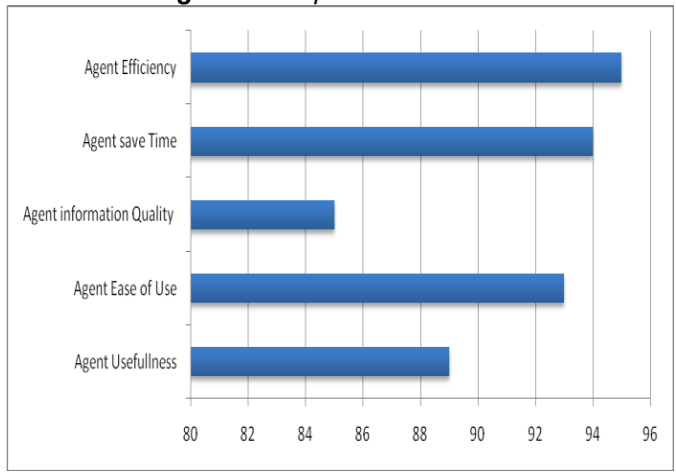

*Figure 6: System's Capability*

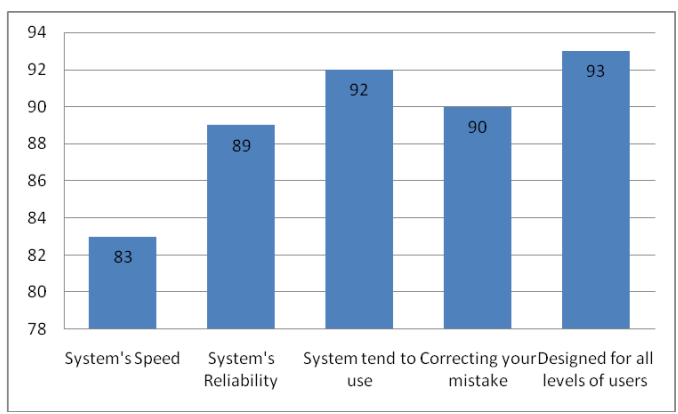

### 7. CONCLUSION AND FUTURE WORK

This paper introduces i-MOBE, a intelligent agent of mobile application for outcome based education in HLI. The main features of OBE intelligent system include academic interface agent, student interface agent, assessment agent and database agent which are helpful to the academicians to evaluate and monitor student performance in a course. The i-MOBE that we developed is considered very important for academicians and students to facilitate them in using for academic purpose in HLI particularly in helping them to monitor the performance in teaching and learning (T&L) and also provide interface for interacting with existing other systems. The conceptual design and system architecture described in this study are successfully translated into functional system as well as the system configuration in helping the academicians to use in their T&L toward effective and efficiency. This study is not without its limitations. First, due to time and resource constraint, the sample size of this study was restricted to a single university. Second, the findings can't be generalized in Malaysia to the whole population of academics may be limited. Therefore, future research can be examine how academician and student could be communicated remotely without face to face communication. Secondly, They are also considered not only within institution but together with other institutions or organizations called Multi- Institution.